\documentclass{article}
\usepackage{graphics}

\makeatletter

\newcommand{\LyX}{L\kern-.1667em\lower.25em\hbox{Y}\kern-.125emX\@}

\makeatother

\begin{document}

\title{Dynamics of Cosmic Necklaces}

\author{Xavier Siemens$^a$, Xavier Martin$^b$ and Ken D. Olum$^a$}

\maketitle
{\par\centering \emph{$^a$Institute of Cosmology, Department of Physics 
and Astronomy,Tufts University, Medford MA 02155, USA}\par}

{\par\centering \emph{$^b$Dpto. de F\'{\i}sica, CINVESTAV-I.P.N. A.P. 
14-74-, 07000 Mexico, D.F., Mexico}\par}

\begin{abstract}
We perform numerical simulations of cosmic necklaces (systems of
monopoles connected to two strings each) and investigate the conditions
under which monopoles annihilate.  When the total monopole energy is
large compared to the string energy, we find that the string motion is
no longer periodic, and thus the strings will be chopped up by self
intersection.  When the total monopole energy is much smaller than the
string energy, the string motion is periodic, but that of the
monopoles is not, and thus the monopoles travel along the string and
annihilate with each other.
\end{abstract}

\section{Introduction}

Topological defects are a natural consequence of phase transitions in the early
universe \cite{1} and are as such predicted by a number of Grand Unified Theories.
They have received much attention in the past two decades, having been considered
promising candidates for a number of interesting cosmological phenomena including
gravity waves, baryon asymmetry, density perturbations, non-gaussianity
in the cosmic microwave background and ultra-high energy cosmic rays. For a
review see \cite{2}. 

Topological defects can occur when some high energy particle physics
symmetry group is broken to a smaller symmetry group. They are formed
by the Kibble mechanism, in which regions of space that are
uncorrelated (at a minimum, those that are outside the causal horizon)
must independently choose vacuum states.  The type of defect formed
depends on the topology of the manifold of equivalent vacuum states.

Cosmic necklaces, in particular, are hybrid topological defects that can be
produced in the sequence of phase transitions 
\[
G\stackrel{\eta _{m}}{\longrightarrow }H\times U(1)\stackrel{\eta _{s}}{\longrightarrow }H\times Z_{2}\,,\]
 where \( G \) is a semi-simple group. In the first phase transition, at an
energy \( \eta _{m} \), monopoles of mass \( m\sim \eta _{m}/e \) are produced.
The second transition, occuring at energy \( \eta _{s} \), traps the magnetic
flux into two strings of mass per unit length \( \mu \sim \eta ^{2}_{s} \)
connecting monopoles to anti-monopoles. The resulting system is a network of
long strings and loops with the monopoles playing the role of beads. 

Previous work by Berezinsky and Vilenkin \cite{3} was done with
the idea of explaining the origin of the highest energy cosmic rays. The
rather complicated underlying dynamics were, for good reason, largely ignored.
They found that if one started with a low enough density of monopoles, such
that one could approximate the evolution of the system using simple string 
evolution,
and more importantly, if one could disregard the effects of
monopole-antimonopole
annihilation, the density of monopoles on the string would naturally increase
to the point where the approximation of simple string evolution would break
down. The monopole density could also be increased with a sufficiently long 
damping era following the formation of the network. 
Furthermore, they noted that a large enough density of monopoles would make 
loop
motion non-periodic and therefore loop fragmentation very efficient. Thus, in
their scenario, the network consits of a long string with monopoles in the 
high 
monopole density regime. The string intercommutes to form loops that rapidly 
fragment and yield cosmic rays by the annihilation of the monopoles.  
The estimated cosmic ray energies and fluxes produced by this model 
seem to fit current observations. They did leave, however,
the detailed analysis of the evolution of these systems to numerical
simulations and, in particular, the verification that
monopole-antimonopole annihilation can initially be disregarded.

In this work we make a first attempt at understanding the dynamics of necklaces
by numerically evolving single loops.

In the next section we review general string motion for completeness 
and derive useful results used in the rest of our paper.  In the third section
we look at monopole motion on cosmic necklaces when the energy of the
monopoles is comparable to the energy in the string. In the fourth
section we examine dynamics and monopole annihilations in the limit
where the monopole energy is low compared to the string energy. We
summarise and conclude in the fifth section.

\section{Review of String Motion}

When the typical length scale of a cosmic string is much larger than its thickness,
\( \delta _{s}\sim \eta ^{-1}_{s} \), and when there are no long-range interactions between
different string segments, as is the case for gauge strings, the string can
be accurately modeled by a one dimensional object. Such an object sweeps out
a two dimensional surface referred to as the string world-sheet. This surface
can be described by a function of two parameters, a timelike parameter $t$ which can be 
identified with the time coordinate, and a spacelike parameter $\sigma$,

\[
x^{\mu }=x^{\mu }(t,\sigma ).\]

The infinitesimal line element in Minkowski space-time with metric \( \eta _{\mu
\nu }={\rm diag} (1,-1,-1,-1) \)
is

\[
ds^{2}=\eta _{\mu \nu }dx^{\mu }dx^{\nu }=\eta _{\mu \nu }x^{\mu }_{,a}x_{,b}^{\nu }d\xi ^{a}d\xi ^{b},\]
 where \( a=0,1 \) labels the internal parameters, \( t=\xi ^{0} \), \( \sigma =\xi ^{1} \) and \( x_{,a}^{\mu }=\partial x^{\mu }/\partial \xi ^{a} \).
One can then write the induced metric on the world-sheet of the string as
\begin{equation}
\label{inducedmetric}
\gamma _{ab}=\eta _{\mu \nu }x^{\mu }_{,a}x_{,b}^{\nu }\,.
\end{equation}

For an infinitely thin string we can use the Nambu-Goto action.  It is
proportional to the invariant area swept by the string,

\begin{equation}
\label{stringaction}
S_{s}=-\mu \int dA=-\mu \int d^{2}\xi \sqrt{-\gamma },
\end{equation}
 where $\gamma=det(\gamma_{ab})$ and \( \mu  \) is the mass per unit length of the string. In geometrical
terms this is the action for a 1+1 dimensional space-time with a cosmological
constant \( \mu  \). Its variation with respect to $x^\mu$ gives the equation
\begin{equation}
\label{dstringaction}
\delta S_{s}=-\mu \int \partial _{a}(\sqrt{-\gamma }\gamma ^{ab}x_{,b}^{\mu }\delta x_{\mu })d^{2}\xi +\mu \int \partial _{a}(\sqrt{-\gamma }\gamma ^{ab}x_{,b}^{\mu })\delta x_{\mu }d^{2}\xi =0.
\end{equation}

The first term in this equation is an integral of a total divergence that can be
turned into boundary terms at the ends of the string. For ordinary strings 
these terms vanish because of the absence of boundaries. 
In the case of cosmic necklaces however, monopoles live on the boundaries
of strings and so these turn out to be the most interesting
terms. The second term gives the usual equations of motion for the
Nambu-Goto string.  If we work, as usual, in the conformal gauge,

\begin{equation}
\label{gauge1}
{\bf x}'(\sigma ,t)\cdot {\bf \dot{x}}(\sigma ,t)=0
\end{equation}

\begin{equation}
\label{gauge2}
{\bf x}'^{2}(\sigma ,t)+{\bf \dot{x}}^{2}(\sigma ,t)=1,
\end{equation}
and choose $t = x^0$, the equation of motion is

\begin{equation}
\label{waveeq}
{\bf x}''(\sigma ,t)={\bf \ddot{x}}(\sigma ,t)\,.
\end{equation}
Primes and dots denote partial derivatives with respect to \( \sigma  \)
and \( t \) respectively. This equation can be readily solved using

\begin{equation}
\label{xeqapb}
{\bf x}(\sigma ,t)=\frac{1}{2}[{\bf a}(\sigma -t)+{\bf b}(\sigma +t)],
\end{equation}
 then

\begin{eqnarray}
\label{xpxdab}
{\bf x}'(\sigma ,t)&=&\frac{1}{2}[{\bf a}'(\sigma -t)+{\bf b}'(\sigma
+t)]\\
{\bf \dot{x}}(\sigma ,t)&=&\frac{1}{2}[-{\bf a}'(\sigma -t)+{\bf b}'(\sigma +t)]
\end{eqnarray}
and the constraints coming from (\ref{gauge1},\ref{gauge2}) become

\begin{equation}
\label{constr}
{\bf a}'^{2}(\sigma -t)={\bf b}'^{2}(\sigma +t)=1.
\end{equation}

The functions \( {\bf a}(\sigma -t) \) and \( {\bf b}(\sigma +t) \), often referred
to as right- and left-movers or halves of the string, are constant along the
lines \( \sigma =t \) and \( \sigma =-t \) respectively on the string world-sheet.

\section{Monopole Motion: The Massive Case}

\subsection{Equations of Motion}

Assuming we can treat the monopoles as point particles living on the string and there
are no unconfined magnetic charges,
the motion for a monopole attached to two strings can be described by the action
\[
S=-m\int ds-\mu \sum ^{2}_{i=1}\int ^{\sigma _{i+}(t)}_{\sigma
_{i-}(t)}dA_{i}\,.\]
\begin{figure}
{\par\centering\resizebox*{0.8\textwidth}{0.3\textheight}{\includegraphics{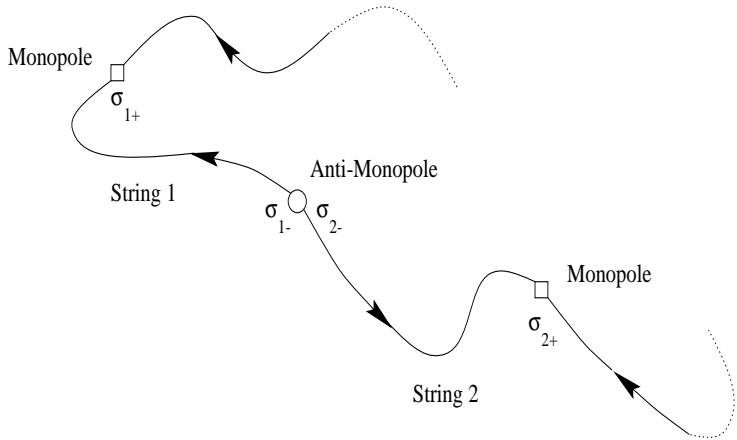}} \par}

\caption{An anti-monopole attached to two monopoles. The string sources (\protect\( \sigma _{i-}\protect \))
are at the anti-monopoles and the ends (\protect\( \sigma _{i+}\protect \))
at monopoles.} \label{fig:string}
\end{figure}%
The first term is the standard action for a relativistic particle of mass $m$, the monopole, \( \mu  \) is the mass per unit length of the string and
\( A_{i} \) are the areas swept by each of the strings attached to the
monopole. The integral over the string world-sheet is bounded by the parameters
\( \sigma _{i-}(t) \) and \( \sigma _{i+}(t) \), the monopole parametric 
positions on
the $i$th string. In our convention the string sources are at anti-monopoles and the
ends at monopoles (see Fig.\ \ref{fig:string}). Here the number of strings attached
is just two but our action can clearly be generalized to any number of strings
by adding new string terms in the sum over $i$.
From (\ref{dstringaction}) one can see the variation of the string part
of the action yields the usual Nambu-Goto term and a total divergence. This
divergence can be turned into two integrals along the world-lines
of the monopoles bounding the string by the 1+1 dimensional analog of Gauss'
theorem \cite{5,4},

\[
\int \partial _{a}(\sqrt{-\gamma }\gamma ^{ab}x_{,b}^{\mu }\delta x_{\mu })d^{2}\xi =-\int \lambda ^-_{a}\gamma ^{ab}x_{,b}^{\mu }\delta x^-_{\mu }ds^--\int \lambda ^+_{a}\gamma ^{ab}x_{,b}^{\mu }\delta x^+_{\mu }ds^+,\]
where the superscripts \( + \) and \( - \) refer to monopoles and
anti-monopoles respectively. The unit vector \( \lambda _{a} \) is orthogonal to the
world-line of the monopole and points into the string world-sheet. In
external coordinates it can be written

\[
\lambda ^{\mu }(t)=\lambda _{a}\gamma ^{ab}x_{,b}^{\mu }=\pm \gamma _{m}(t)[\dot{\sigma }(t)\dot{x}^{\mu }(t,\sigma (t))+x'^{\mu }(t,\sigma (t))],\]
where \( \gamma _{m}(t)=(1-{\bf \dot{x}}_{m}^{2}(t))^{-1/2} \) is the Lorentz
factor of the monopoles, and the upper sign is for anti-monopoles and
the lower for monopoles. 
These boundary terms can be absorbed into the monopole action to give the 
equations of motion for the monopoles \cite{4,5},
\begin{equation}
\label{4accel}
m\frac{d^{2}x^{\nu }}{ds^{2}}=\mu \sum ^{2}_{i=1}\lambda _{i}^{\nu }\,,
\end{equation}
 where the \( i=1,2 \) labels the strings to which the monopole is attached. The
\( 0 \)-component of this equation,

\begin{equation}
\label{encons}
m\dot{\gamma }_{m}(t)=\mu \sum ^{2}_{i=1}\dot{\sigma }_{i}(t)\,,
\end{equation}
where \( \dot{\sigma }_{i}(t) \) is the rate of change
of the position of the monopole on the \( i \)th string, is the equation for
energy conservation. The monopoles can create string and lose kinetic energy
in the process, or annihilate string and gain kinetic energy. The spatial part
in turn can be put in the form

\begin{equation}
\label{accel}
{\bf \ddot{x}}_{m}(t)=\pm \frac{\mu }{m}\gamma ^{-3}_{m}(t)\sum ^{2}_{i=1}\frac{{\bf x}_{i}'(\sigma _{i}(t),t)}{\left| {\bf x}_{i}'(\sigma _{i}(t),t)\right| ^{2}}=\frac{\mu }{m}\gamma ^{-3}_{m}(t)\sum ^{2}_{i=1}\gamma _{i}(t)\hat{n}
\end{equation}
 where \( \gamma _{i}(t) \) is the gamma factor of the string at the position of the monopole, \( {\bf x}_{i}'(\sigma _{i}(t),t) \) is the tangent
vector of the string at the monopole, and \( \hat{n} \) is a unit vector tangent
to the string pointing inwards.
A similar result was obtained in \cite{4} for the case
of strings bounded by monopoles. 

The rate of change of monopole position along the string is related to the energy
balance of the system. A more useful expression than (\ref{encons}) can
be obtained taking the time derivative of the constraint that the monopole must
lie on the string, \( {\bf x}_{m}(t)={\bf x}_{i}(\sigma (t),t) \),

\begin{equation}
\label{xmdot1}
{\bf \dot{x}}_{m}(t)=\frac{d{\bf x}_{i}(\sigma
_{i}(t),t)}{dt}=\dot{\sigma}_{i}(t){\bf x}_{i}'(\sigma _{i}(t),t)+{\bf
\dot{x}}_{i}(\sigma (t),t)\,.
\end{equation}
Using the constraints (\ref{gauge1}) and (\ref{gauge2}), (\ref{xmdot1}) can be re-written as

\begin{equation}
\label{sigmadot}
\dot{\sigma }_{i}(t)=\frac{{\bf \dot{x}}_{m}(t)\cdot {\bf x}_{i}'(\sigma _{i}(t),t)}{\left| {\bf x}_{i}'(\sigma _{i}(t),t)\right| ^{2}}\,,
\end{equation}
 which expresses the rate of change of the monopole parametric position 
along each of the
two strings individually. 

The equations of motion for the monopoles, (\ref{accel}) and
(\ref{sigmadot}), and for the string, (\ref{gauge1}),
(\ref{gauge2}) and  (\ref{waveeq}), could be solved to give the motion of 
our system.
Except for very special initial conditions, however, the task of finding 
analytic
solutions to this system of equations is hopeless, so we resort to numerical
simulation.

\subsection{The Numerical Algorithm}

Although we cannot solve the entire system analytically, we do have
an explicit solution for the interior of the string, (\ref{xeqapb}).  
We can use this solution to give the string
motion, taking as input the motion of the monopoles at the ends of the
string, which we will simulate.  The effect of the string will be to
take excitations from the monopole (or anti-monopole) at one end and
transport them to affect the anti-monopole (or monopole) at the
other. We proceed as follows.

We consider, for simplicity, only a monopole attached to an anti-monopole by a
string, and ignore the effects of the strings attached on the other side of 
either
defect. Since the monopole is constrained to live on the boundary \( \sigma _{i}(t) \)
of the  string, its position in terms of the right- and left-movers
can be written
\begin{equation}
\label{xab}
{\bf x}_{m}(t)={\bf x}_{i}(\sigma _{i}(t),t)=\frac{1}{2}\left[ {\bf a}_{i}(\sigma _{i}(t)-t)+{\bf b}_{i}(\sigma _{i}(t)+t)\right] .
\end{equation}
 Its velocity is then given by

\begin{equation}
\label{xdxspxsd}
{\bf \dot{x}}_{m}(t)=\frac{d{\bf x}_{i}(\sigma _{i}(t),t)}{dt}=\dot{\sigma }_{i}(t){\bf x}_{i}'(\sigma _{i}(t),t)+{\bf \dot{x}}_{i}(\sigma _{i}(t),t).
\end{equation}
 which can be expressed in terms of the right- and left-movers,

\begin{equation}
\label{xdapbp}
{\bf \dot{x}}_{m}(t)=\frac{1}{2}\big\{[\dot{\sigma }_{i}(t)-1]{\bf
a}_{i}'(\sigma _{i}(t)-t)+[\dot{\sigma }_{i}(t)+1]{\bf b}_{i}'(\sigma
_{i}(t)+t)\big\}\,.
\end{equation}

For the anti-monopole, we can interpret this expression as giving the value of \(
{\bf a}_{i}' \), the excitations that travel away from the defect,
in terms of \( {\bf \dot{x}}_{m}\), the monopole velocity,
and \( {\bf b}_{i}' \), the excitations traveling
toward the defect.  Similarly for
the monopole, (\ref{xdapbp}) gives \( {\bf b}_{i}' \) in terms of \( {\bf \dot{x}}_{m} \) and \( {\bf
a}_{i}' \). The outgoing
excitations \( {\bf a}_{i}'(\sigma _{i}(t)-t) \) and \( {\bf
b}_{i}'(\sigma _{i}(t)+t) \) then propagate along the string segment into the
future, and eventually reach the other end (see Fig.\ 2).
\begin{figure}
{\par\centering \resizebox*{0.7\textwidth}{0.3\textheight}{\includegraphics{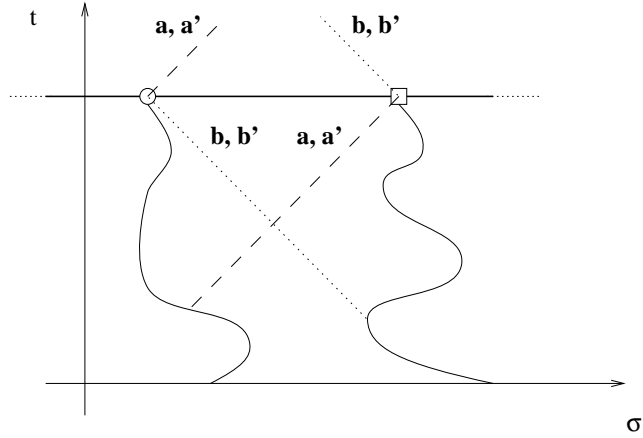}}\par}

\caption{The excitations on the string travel from monopole to anti-monopole and vice-versa,
getting transformed with each bounce.} \label{fig:asandbs}
\end{figure}

The equations for the system, after some algebra, can be cast in the form

\begin{eqnarray}
\label{sdM}
\dot{\sigma }_{i}(t)&=&\frac{{\bf \dot{x}}^{2}_{m}(t)+{\bf \dot{x}}_{m}(t)\cdot {\bf a}_{i}'(\sigma _{i}(t)-t)}{{\bf \dot{x}}_{m}(t)\cdot {\bf a}_{i}'(\sigma _{i}(t)-t)+1}\\
\label{bpM}
{\bf b}_{i}'&=&\frac{2{\bf \dot{x}}_{m}-[\dot{\sigma }_{i}(t)-1]{\bf a}_{i}'(\sigma _{i}(t)-t)}{\dot{\sigma }_{i}(t)+1}
\end{eqnarray}
 for the monopole, and

\begin{eqnarray}
\label{sdAM}
\dot{\sigma }_{i}(t)&=&\frac{{\bf \dot{x}}^{2}_{m}(t)-{\bf \dot{x}}_{m}(t)\cdot {\bf b}_{i}'(\sigma _{i}(t)+t)}{{\bf \dot{x}}_{m}(t)\cdot {\bf b}_{i}'(\sigma _{i}(t)+t)-1}\\
\label{apAM}
{\bf a}_{i}'&=&\frac{2{\bf \dot{x}}_{m}-[\dot{\sigma }_{i}(t)+1]{\bf b}_{i}'(\sigma _{i}(t)+t)}{\dot{\sigma }_{i}(t)-1}
\end{eqnarray}
 for the anti-monopole. These equations, along with
 (\ref{accel}) and
\[
{\bf x}_{i}'(\sigma _{i}(t),t)=\frac{1}{2}[{\bf a}_{i}'(\sigma _{i}(t)-t)+{\bf b}_{i}'(\sigma _{i}(t)+t)]\]
 can be used to solve for the motion of any system of this type.

In a typical time-step of our code at, say, an anti-monopole, we have its velocity,
\( {\bf \dot{x}}_{m}(t) \), and the values of \( \sigma _{1}(t) \) and
\( \sigma _{2}(t) \), its parametric positions
on the strings. We also have a sequence of values of \( {\bf b}_{i}'
\) emitted at past times from the other end of the string (see Fig.\ \ref{fig:asandbs}).  By
interpolation, we can find \( {\bf b}_{i}' (\sigma_i (t) +t)\)
 and use it to compute \( \dot{\sigma }_{i}(t) \)
and \( {\bf a}_{i}' \) using (\ref{sdAM}) and  (\ref{apAM}). If we do this
for both strings we can compute the acceleration according to (\ref{accel}).
We can then use the acceleration and the \( \dot{\sigma }_{i}(t) \) we just
calculated to compute the velocity and the \( \sigma _{i} \) at the next time-step
(with an appropriate finite differencing scheme). This is what we started with
and therefore all we need to continue onto the next timestep. It is, of course,
necessary to store enough of the \( {\bf a}_{i}' \) and \( {\bf b}_{i}' \)
to ensure we can perform the appropriate interpolations at every time-step. We
use energy conservation, (\ref{encons}), as an independent check on the accuracy
of our code (for example see Fig.\ \ref{energyplot}).

The reward for our efforts is that in a numerical computation we can ignore
the presence of the strings entirely and simply store the outward going string
excitations in the past of the monopole and anti-monopole world-lines to be
used when needed. 

\begin{figure}
{\par\centering \resizebox*{0.7\textwidth}{0.3\textheight}{\includegraphics{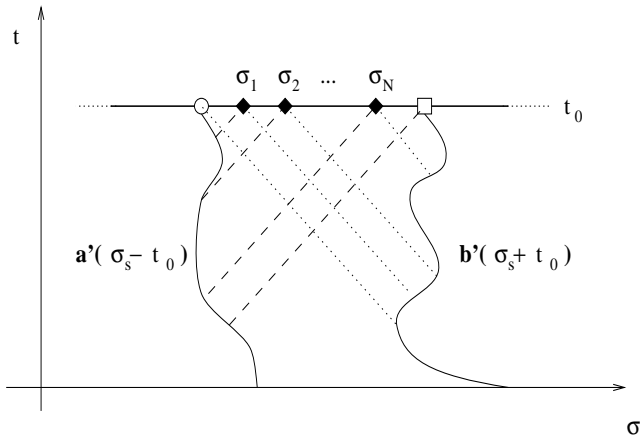}} \par}

\caption{The position of the string can be readily reconstructed at any time \protect\( t_{0}\protect \)
by starting one end of the string, finding a value \protect\( \sigma _{s}\protect \)
on the string, interpolating for the values of \protect\( {\bf a}'\protect \)
and \protect\( {\bf b}'\protect \) using \protect\( \sigma _{s}-t_{0}\protect \)
and \protect\( \sigma _{s}+t_{0}\protect \) respectively and then integrating
to find \protect\( {\bf a}\protect \) and \protect\( {\bf b}\protect
\).}
\label{fig:reconstruction}
\end{figure}

If one is interested in the string, however, its position or velocity can be
readily re-constructed: Starting at one end of the string,  one can find
the values of \( \sigma _{s}-t \) and \( \sigma _{s}+t \) that correspond
to a certain point \( \sigma _{s} \) on the string, look up $\bf
a'$ and $\bf b'$, and integrate to get $\bf a$ and $\bf b$, and thus
the string position (see
Fig.\ \ref{fig:reconstruction}). 

\subsection{Results}

A fundamental feature of these systems, first pointed out by Berezinsky and Vilenkin 
\cite{3}, is the existence of the dimensionless parameter \( \mu L/Nm \),
the ratio of the total string energy to the total monopole energy. Here, \( \mu  \)
is the mass per unit length of the string, \( L \) the invariant length, \( N \)
the number of monopoles and \( m \) the monopole rest mass. This parameter
defines an equivalence class of loops in the sense that loops with the same
geometrical form (shape) and the same value of that parameter will evolve in
an analogous way. One can easily see this equivalence by looking at the equations
of motion for the monopole (\ref{4accel}). This is, in fact, a straightforward
generalization to our case of the usual conformal invariance of cosmic strings.
For example, if one changes the lengths of string on a loop, \( L\rightarrow \lambda L \),
and the monopole masses, \( m\rightarrow \lambda m \), the value of \( \mu L/Nm \)
remains constant and the evolution will be the same in the sense that after
a time \( \lambda t \) the larger loop will look the same as the smaller one
at time \( t \), remaining, of course, \( \lambda  \) times bigger. 

We have evolved a considerable number of different loop configurations with
values of the dimensionless parameter \( \mu L/Nm\sim 1 \). As was anticipated,
one of the most important effects is the non-periodicity of the solutions which
has rather notable consequences, making the evolution of these systems very
different from that of ordinary cosmic strings. In particular, the non-periodic
nature of the solutions allows the system to explore the phase space. Since
there are many more ways in which the string can be wiggly than straight it
ends up being mostly wiggly; this is a vastly different situation from that
of regular cosmic strings where the loops can get trapped in smooth low entropy 
configurations and never explore the phase space because of the periodicity of 
their motion. 

\begin{figure}
{\par\centering \resizebox*{0.8\textwidth}{0.4\textheight}{\includegraphics{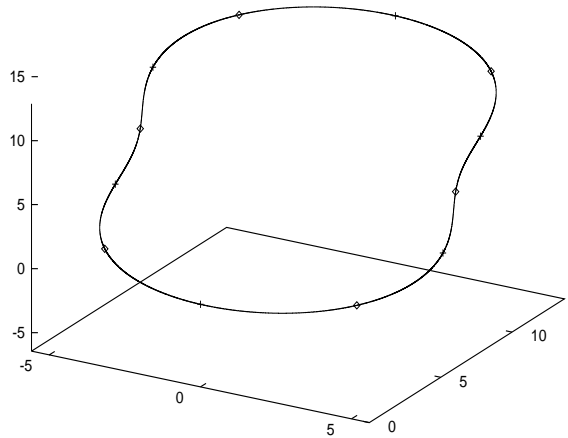}} \par}

\caption{Plot of initial loop with 12 monopoles. It is a Kibble-Turok loop \cite{6}
with parameter values of \protect\( \kappa =.7\protect \) and \protect\( \phi =\pi /7\protect \).
The value for the dimensionless parameter is \protect\( \mu L/Nm=5\protect \)
and the resolution used is \protect\( \delta t=10^{-3}\protect \).\label{straightloop}}
\end{figure}
\begin{figure}
{\par\centering \resizebox*{0.8\textwidth}{0.4\textheight}{\includegraphics{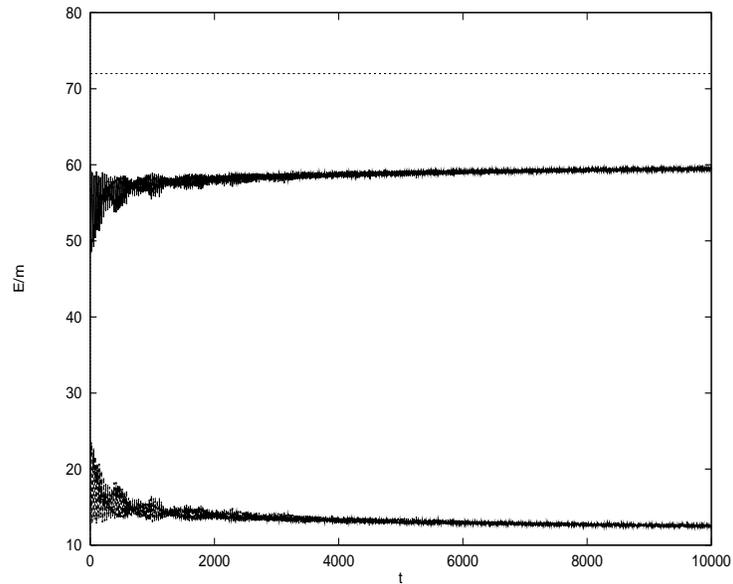}} \par}

\caption{The energy in units of the monopole mass for the necklace loop shown 
in Fig.\ \ref{straightloop}, 
as a function of time. The bottom line is the monopole energy, the second line
is the string energy and the top line is the total energy. The system has been
evolved for a time \protect\( t=10000=P\delta \protect \), where \protect\( P\protect \)
is the number of timesteps in the code. During this time it executes approximately
300 `oscillations' where what we mean by `oscillations' is number of local maxima
of the monopole (or string) energy. The energy is very well conserved: At the
end of our simulation \protect\( \Delta E/E\sim 10^{-4}\protect \).\label{energyplot}}
\end{figure}

\begin{figure}
{\par\centering \resizebox*{0.8\textwidth}{0.4\textheight}{\includegraphics{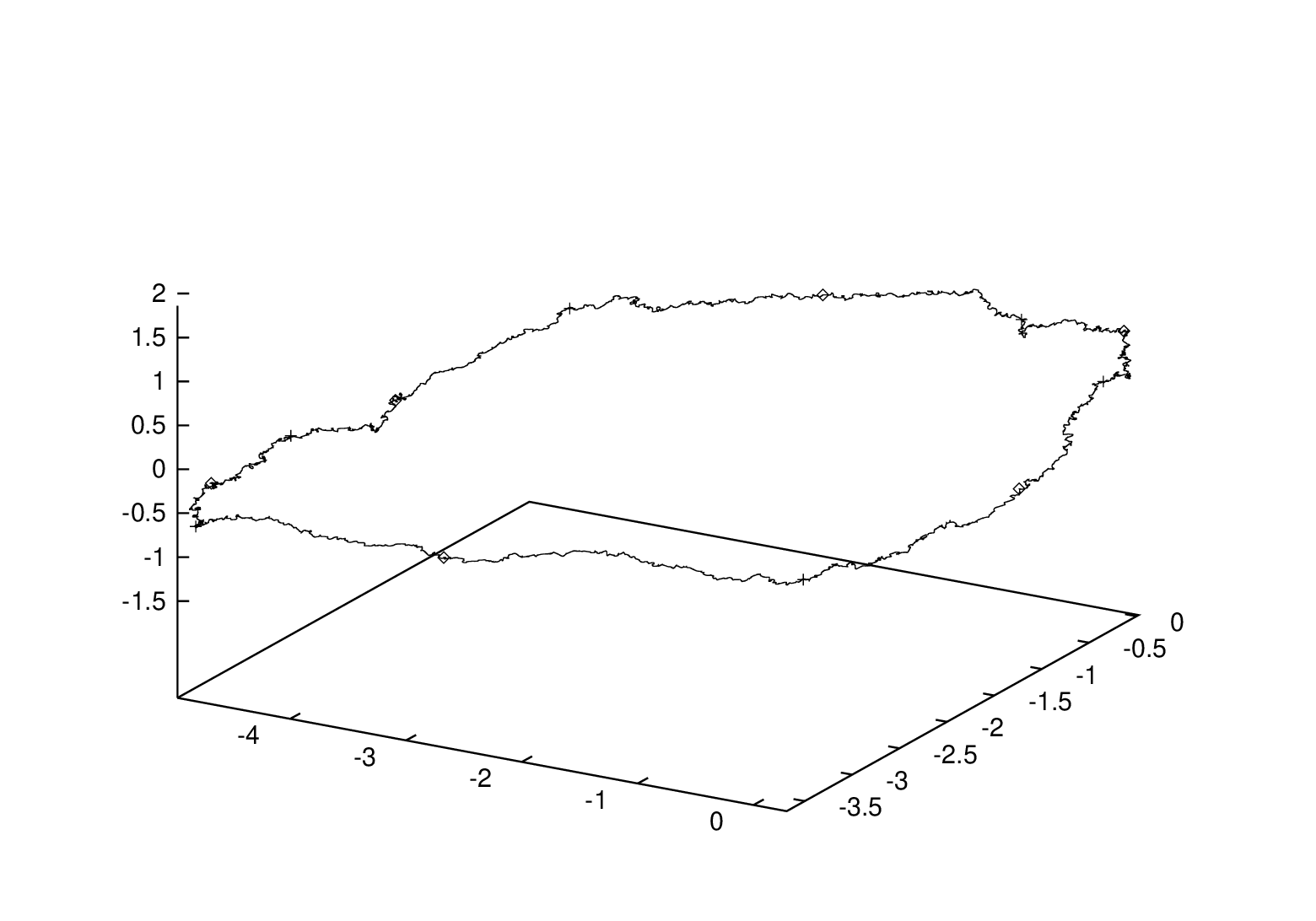}} \par}

\caption{Plot of the same loop as in Fig.\ \ref{straightloop} at \protect\( t=10000\protect \),
after 300 `oscillations'. Notice the size of the loop is considerably smaller: A substantial 
amount of the original string length has formed wiggles.}\label{wigglyloop}
\end{figure}

Here we only show the results for one of the loop configurations we have evolved.
However, the result is typical of all our other simulations and the difference
between the usual cosmic string evolution and cosmic necklace evolution is very
well illustrated. 

Figs.\ \ref{straightloop}, \ref{energyplot} and \ref{wigglyloop} show
the evolution of a necklace using as initial conditions a Kibble-Turok 
loop \cite{6} with values for the loop parameters
of \( \kappa =0.7 \) and \( \phi =\pi /7 \), twelve monopoles and a value for
the dimensionless parameter \( \mu L/Nm=5 \). One can see from Fig.\ \ref{energyplot}
that the monopole and string energies go through a series of maxima and minima.
Bearing in mind that the motion is not truly periodic we shall nevertheless abuse
the language and refer to these as oscillations. 

In the early stages of evolution, when the strings are still quite straight,
the monopoles can easily gain kinetic energy from the strings. However, after
a few oscillations, thirty or so, the strings become so wiggly that the monopoles
can no longer gain significant amounts of kinetic energy from them. While the details of
the mechanism for wiggle formation on the string elude us it is clear that they  
arise from the non-linear back-reaction of the monopoles on the string \ref{bpM}, 
\ref{apAM}.
After the wiggles are formed the energy of the monopoles is mostly just 
their rest mass. We have observed
the behaviour described here in all loops in our simulations with values of
the dimensionless parameter \( \mu L/Nm\sim 1 \). 

Although we have not included self-intersections in our code, it seems 
clear that had we included them the loop would have fragmented well before 
reaching the wiggly state shown in Fig.\ \ref{wigglyloop}. While the 
exact timescale is not obvious the general evolution is unambiguous.
Because of the non-periodicity of necklace 
motion, necklace loops fragment into smaller
daughter loops; some containing monopoles and others not. 
The ones not containing
monopoles evolve like ordinary Nambu-Goto string loops: They self-intersect 
until they find themselves in stable trajectories and then 
decay by gravitational radiation. The ones that do contain monopoles
self-intersect again until only monopole-antimonopole pairs are left on 
each of the
loops. These pairs then radiate small loops and gravitational radiation until
they decay and the monopole and anti-monopole annihilate, along the
lines discussed in \cite{4,7}.

\section{Monopole Motion: The Massless Limit}

The starting regime for the necklace network proposed by Berezinsky and 
Vilenkin 
\cite{3} is the limit where the string energy is much larger than the monopole
energy.  Unless there is a sufficiently long damping era following network 
formation we expect the dynamics to be quite different from what we have 
described in the previous section. 

Because of the equivalence of loops with equal values of the dimensionless 
parameter
\( \mu L/Nm \) the large string energy limit can be thought of as the limit
where the monopole mass \( m \) is negligible compared with the mass per unit
length of the string \( \mu  \). Unfortunately, we can no longer use the 
equations
derived in the previous section because they do not have a well 
defined \( m\rightarrow 0 \)
limit. 

It turns out to be convenient to think of the monopoles as test particles constrained
to live on a 1+1 dimensional dynamical space-time, the string. In this case
the monopoles do not back-react on the string and therefore also do not produce small-scale 
structure
so we expect them to get accelerated by the curvature of the string and eventually
collide and annihilate.

The motion of the string, being unaffected by the presence of the test particles,
is given by some solution of the Nambu-Goto equations of motion. The monopole
motion can in turn be derived by minimising the 1+1 action

\[
S_{m}=-m\int ds=-m\int \sqrt{\gamma _{ab}d\xi ^{a}d\xi ^{b}}\]
 which is the world-line of the monopole on the string world-sheet. One can
re-write this action by explicitly evaluating \( \gamma _{ab} \) in the conformal
gauge, yielding

\[
S_{m}=-m\int \sqrt{(1-{\bf \dot{x}}_{s}^{2})dt^{2}-{\bf x}_{s}'d\sigma ^{2}}=-m\int dt\sqrt{1-{\bf \dot{x}}_{m}^{2}}\]
 where

\[
{\bf \dot{x}}_{m}(t)=\frac{d{\bf x}_{s}(\sigma (t),t)}{dt}=\dot{\sigma }(t){\bf x}_{s}'+{\bf \dot{x}}_{s}\]
 is the monopole velocity. It should be noted that this action could have also
been derived using the constraint \( {\bf x}_{m}(t)={\bf x}_{s}(\sigma (t),t) \)
and taking the action to be the world-line of the monopole in 3+1 Minkowski
space instead of the 1+1 world-sheet of the string. We feel, however, that the
derivation outlined above makes the degrees of freedom of the monopole motion
more manifest. The only degrees of freedom of the monopole are
its parametric position \( \sigma (t) \) and velocity \( \dot{\sigma }(t) \) along the
string. 

This action yields the equation for the motion of the monopole along the string
which, after some algebra, can be put in the form

\begin{equation}
\label{sigmaddot}
\ddot{\sigma }(t)=(\dot{\sigma }(t)-1)^{2}\frac{{\bf x}_{s}'(\sigma ,t)\cdot \left\{ {\bf \ddot{x}}_{s}(\sigma ,t)+\dot{\sigma }(t){\bf \dot{x}}_{s}'(\sigma ,t)\right\} }{\left| {\bf x}_{s}'(\sigma ,t)\right| ^{2}}
\end{equation}

This equation can be readily solved by numerically integrating the velocity
\( \dot{\sigma }(t) \) and position \( \sigma (t) \) along the string of monopoles
and anti-monopoles for a number of different loop trajectories. 

\begin{figure}
{\par\centering \resizebox*{0.8\textwidth}{0.4\textheight}{\includegraphics{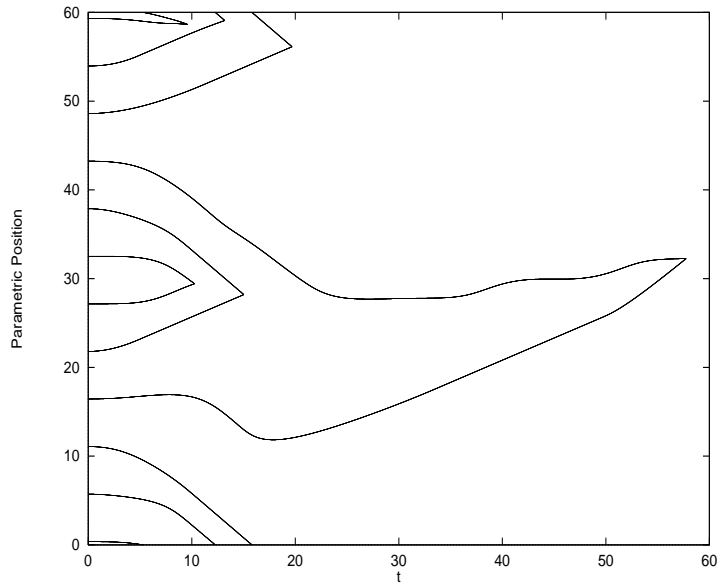}} \par}
\caption{Plot of the parametric position $\sigma (t)$ for twelve monopoles
on a Kibble-Turok \cite{6} loop with a length $L=60$ and values for the 
parameters the same as in Fig.\ \ref{straightloop}. The monopoles' initial 
parametric position is not quite the same as in the previous figure because we want 
to avoid degeneracies arising from the symmetry of loop motion. The string 
is periodic in its length  
and the monopoles that cross the bottom boundary end up at the top.
As can be seen from the figure the monopoles move on the string fairly 
randomly and all of them 
annihilate rather quickly, with the last pair annihilating right 
before the end of the second oscillation period of the loop.}\label{sigmaplot}
\end{figure}

We have again used a wide variety of initial conditions and 
found that, as we expected, in all cases the monopoles annihilate with
the anti-monopoles within a few oscillation periods. The typical behaviour 
is well illustrated in Fig.\  \ref{sigmaplot} which shows the evolution of the 
parametric position of twelve monopoles on a loop identical to the one in 
Fig.\ \ref{straightloop}.  All twelve monopoles annihilate before the end the second 
oscillation period. Although the exact timescale does depend on the 
particular shape of the loop, in no case have we found it to be longer 
than a few oscillation periods of the loop. The reason for this is
that the motion of the string is unaffected by the presence of the monopoles
and is therefore periodic, while the motion of the monopoles given by 
(\ref{sigmaddot}), is not in general
periodic. Being constrained to live on a string, it is inevitable that 
monopoles eventually collide and annihilate.

\section{Concluding Remarks}

We have examined the dynamics of cosmic string loops with monopoles in 
two limits, when the energy in the string is comparable
to the monopole energy and when the string energy is much larger than 
the monopole energy. 

The behaviour in both cases seems to point to the annihilation of the
monopoles in the system. When the monopole energy is very small compared 
with the string energy, the
monopoles on the string collide and annihilate in a few oscillation
periods.  When the monopole energy is large, self intersections chop
the string into small loops which subsequently decay by gravitational
radiation. If the monopole and string scales are not too different,
this process inevitably leads to rapid monopole annihilation. On the 
other hand if
the monopoles are very heavy and the strings very light, loops with
one monopole and one anti-monopole may be long-lived, as in
\cite{7}. Except for this last possibility, though, it would appear
that necklace loops are rather short lived.

However, one could envision an intermediate regime in which the back reaction
of the monopoles on the string is not large enough to produce such a rapid 
process of decay by self intersections but not so small as to decay directly 
by annihilations. Such a regime could significantly increase the life-time 
of loops. Unfortunately, at present, we are not able to faithfully 
simulate such a regime.

Furthermore, in the case when the monopole energy is small compared to the 
string energy, the possibility exists that monopoles may miss each other upon
colliding and avoid annihilation. The likelihood of this would depend on the
cross section for monopole-antimonopole annihilation and the thickness of the
string. If the probability of crossing is large then having crossed each other
they would feel a force towards each other (twice the tension on the string)
and undergo a bouncing process. Such a process could in principle extend the 
lifetime of the monopoles on the string until the strings have radiated 
enough energy to
gravitational waves that the monopole energy becomes comparable to the string
energy. At this point the system would decay by self-intersections,
which could in principle be happening now. Our results therefore indicate that 
if the string energy is dominant when the system is formed and there is not 
sufficient damping to overturn this regime, the scenario for cosmic ray 
production 
proposed in \cite{3} is viable only if the cross section for 
monopole-antimonopole annihilation is sufficiently small. 

The purpose of this work was to study the dynamics of cosmic necklaces to use
this knowledge in the construction of cosmologically interesting models. 
We have found that
we can understand the behaviour of these systems in two different regimes. A
more definite statement regarding the viability of these systems as a source
of ultra-high energy cosmic rays and other cosmological consequences of 
interest depends crucially on the cosmological evolution of the 
dimensionless parameter \( \mu L/Nm \). 
Work in this direction is in progress.

\section{Acknowledgements}

We would like to thank Jose Blanco-Pillado, Christopher Borgers, Carlos Serna,
Alexander Vilenkin and Serge Winitzki for useful discussions. The work of 
K.D.O. was supported in part by the National Science Foundation.


\begin{thebibliography}{}
\bibitem{1}T.W.B. Kibble, J. Phys. A 9 (1976) 1387; Phys. Rep. 67 (1980) 183.
\bibitem{2}A. Vilenkin and E.P.S Shellard, Cosmic Strings and Other Topological Defects (Cambridge University Press, Cambridge, 1994).
\bibitem{3}V. Berezinsky and A. Vilenkin, Phys. Rev. Lett. 79 (1997) 5202.
\bibitem{4}X. Martin and A. Vilenkin, Phys. Rev. D55 (1997) 6054.
\bibitem{5}B. Carter, The Formation and Evolution of Cosmic Strings, eds. G.W. Gibbons, S.W. Hawking and T. Vachaspati (Cambridge University Press, Cambridge, 1990), p. 143.
\bibitem{6}T.W.B. Kibble and N. Turok, Phys. Lett. B 116 (1982) 141.
\bibitem{7}J.~J.~Blanco-Pillado and K.~D.~Olum,
Phys. Rev.  D 60 (1999) 083001.
\end{thebibliography}
\end{document}